\documentclass[useAMS,usenatbib]{mn2e}

\title[Masers with EGOs]
  {Class I methanol masers: Masers with EGOs}
\author[X. Chen et al.]
  {X. Chen,$^1$\thanks{E-mail: chenxi@shao.ac.cn}
  S. P. Ellingsen$^2$
  and Z.-Q. Shen$^1$ \\
  $^1$ Key Laboratory for Research in Galaxies and Cosmology, \\
  \ \ Shanghai Astronomical Observatory, Chinese Academy of Sciences, 80 Nandan Road, Shanghai 200030, China\\
  $^2$School of Mathematics and Physics, University of Tasmania, Hobart, Tasmania, Australia}

\pagerange{\pageref{firstpage}--\pageref{lastpage}} \pubyear{2009}

\def\LaTeX{L\kern-.36em\raise.3ex\hbox{a}\kern-.15em
    T\kern-.1667em\lower.7ex\hbox{E}\kern-.125emX}

\begin{document}

\label{firstpage}

\maketitle

\begin{abstract}
 We have compared the results of a number of published class~I
methanol maser surveys with the catalogue of high-mass outflow
candidates identified from the GLIMPSE survey (known as extended
green objects or EGOs). We find class~I methanol masers associated
with approximately two-thirds of EGOs. Although the association
between outflows and class~I methanol masers has long been
postulated on the basis of detailed studies of a small number of
sources, this result demonstrates the relationship for the first
time on a statistical basis. Despite the publication of a number
of searches for class~I methanol masers, a close physical
association with another astrophysical object which could be
targeted for the search is still lacking. The close association
between class~I methanol masers and EGOs therefore provides a
large catalogue of candidate sources, most of which have not
previously been searched for class~I methanol masers. Interstellar
masers and outflows have both been proposed to trace an
evolutionary sequence for high-mass star formation, therefore a
better understanding of the relationship between class~I methanol
masers and outflow offers the potential for comparison and
amalgamation of these two evolutionary sequences.
\end{abstract}


\section{Introduction}

The brightest Galactic interstellar masers are associated with
regions of high-mass star formation, many of which show emission
from multiple maser species or transitions. The most commonly
observed and strongest masers in star formation regions are 22-GHz
H$_{2}$O masers, main-line OH masers, class II methanol masers
(particularly the 6.7- and 12.2-GHz transitions) and class I
methanol masers. Of these different types of masers, the least
well studied and understood are the class~I methanol masers.

The two classes of methanol maser were initially proposed on an
empirical basis by \citet{BM88} and \citet{M91}. They observed
that some transitions were closely associated with OH and water
masers and bright infrared emission (class II methanol masers),
while other transitions were typically offset from such
infrared-bright areas of star formation regions by up to 1 pc
(class I methanol masers). The best-studied methanol masers are
the 6.7-GHz transition (class II) which has been the subject of
numerous large-scale searches since its discovery and there are
now more than 800 sources known in the Galaxy \citep{PMB05,G+09}.
The class II methanol masers are only associated with high-mass
star formation regions \citep{MENB03} and also trace an early
evolutionary stage, as evidenced by their association with
infrared dark clouds (IRDC) \citep{E06} and millimeter and
submillimeter dust continuum emission \citep{P+02,W+03}. Some of
the strongest class~I methanol maser sources were naturally the
first to be studied in detail. Observations of DR21 \citep{PM90}
and Orion KL \citep{J+97} found the masers were located at the
interface between molecular outflows and the parent cloud. These
sources also typically showed either no, or very weak emission
from the strong class II maser transitions and it was commonly
thought that class I and class II methanol masers favored very
different environments. This view was supported by early
theoretical models of methanol masers which showed strong maser
emission from class I transitions when collisional processes are
dominant and strong masers in the class II transitions when
radiative processes dominate \citep{C+92}.

The first indications that the relationship between the two
classes of methanol masers might be more complex came when
\citet{S+94} detected 44-GHz class I methanol masers towards a
large number of class II maser targets. Subsequent observations of
the 95-GHz class I methanol masers by \citet{V+00} and \citet{E05}
confirmed this. The nature of the relationship between the two
classes of methanol masers remains uncertain. The two classes are
``associated'' in the sense that a search for class~I methanol
masers towards a known class~II site will often find a source
within 30 ~arcseconds, however, whether there is a direct
relationship between the two classes, a ``physical association",
is not yet clear (hereafter, the terms association and physical
association should be understood to take the meaning outlined
here). More sophisticated modelling has since found that in some
cases weak class II methanol masers can be associated with strong
class~I masers and vice versa, as may be the case in Orion
\citep{V+05}. Studies of class I methanol masers have lagged those
of the other common species in star formation regions for two
reasons, one is that there are no strong, common class I
transitions at low frequencies. The other reason is that to date,
there has been no close association of class I masers with other
astrophysical objects that could be used for targeted searches.
\citet{E05} searched for 95-GHz class~I methanol masers towards a
statistically complete sample of 6.7-GHz class II methanol masers
and achieved a detection rate of approximately 40\%.

Those class~I methanol masers which have been studied at high
resolution typically are found at the interface between an outflow
and the parent molecular cloud \citep[e.g.][]{V+06}. However, the
close association between outflows and class I masers has only
been established in a small number of sources. One of the problems
has been in finding appropriate outflow tracers. Shocked H$_2$
exhibits characteristic narrow-band emission at 2.12~$\mu$m,
however, \citet{V+06} found it to be associated with only some of
the class I masers in {\em IRAS}16547--4247 (this source is also
known as G343.12-0.06 and it is listed in Table 3 under that name
in this work). It is not clear if this is due to variable
extinction across the source, or whether some of the masers are
associated with outflows which are not energetic enough to produce
2.12~$\mu$m H$_2$ emission. Molecular tracers such as SiO are
often associated with bipolar outflows in both high- and low-mass
star formation regions, however, in the cluster environment
multiple outflows are frequently observed \citep{B+02,Q+08,dB+09}
and while some of these may be associated with class I masers, in
many cases they are not. \citet{VL07} have compiled a catalogue of
160 class~I methanol maser sources (incorporating all the
published surveys for class I sources) and undertaken a
statistical analysis of their associations. They found that 72\%
of class~I methanol maser sources are associated with class~II
methanol masers, 91\% with H$_2$O masers. Somewhat surprisingly
only 25\% were found to be associated with bipolar outflows
(within 2\arcmin).

Recently \citet{C+08} have identified a new and powerful outflow
tracer for high-mass star formation regions - extended emission in
{\em Spitzer} IRAC images seen predominantly in the 4.5~$\mu$m
band. The strong, extended emission in this band is thought to be
produced by shock-excited H$_2$ and CO and \citet{C+08} have
demonstrated that these sources, (dubbed extended green objects or
EGOs, after the common color-coding of the 4.5~$\mu$m band as
green in three-color images) are in many cases associated with
IRDC and class II methanol masers.

\citet{C+08} found that 6.7-GHz class II methanol masers are
associated with 73\% of ``likely'' massive young stellar object
(MYSO) outflow candidate EGOs and 27\% of ``possible'' MYSO
outflow candidate EGOs. However, given that EGOs signpost outflows
and class I masers are also known to trace outflows it seems
plausible that there may be a relationship between these two
astrophysical phenomena. Section 2 details a comparison of class I
methanol maser observations from the literature with the EGO
catalogue of \citet{C+08}, while \S~3 discusses the significance
of the findings and possible future observations to further
elucidate the nature of the relationship.

\section{EGOs, outflows and class I methanol masers}

To investigate if there is any relationship between EGOs and
class~I methanol masers we have compared the results of four maser
searches with the EGO list compiled by \citet{C+08}.
\citeauthor{C+08} list a total of 302 EGOs, of which 137
(including 4 EGOs showing distinct emission from their central
objects listed in Table 5 of Cyganowski et al. 2008) are
classified as ``likely'' and 165 as ``possible'' outflow sources,
we have combined these into a single EGO sample. The four class~I
maser surveys include searches for the 44-GHz $7_0$--$6_{1}$
A$^{+}$ transition by \citet{S+94} and \citet{KHA04} and searches
for the 95-GHz $8_0$--$7_{1}$ A$^{+}$ transition by \citet{V+00}
and \citet{E05}. We selected these four surveys for our
statistical analysis because they include a large fraction
(135/160) of the known class~I maser sources \citep{VL07}, and
they also list non-detections from their searches which allows a
more detailed analysis to be undertaken. The results of these four
surveys are summarised in Table~1. Examination of Table 1 reveals
that the overall detection rate for the Slysh et al. (1994) survey
is significantly lower than the other three surveys, while it is
significantly higher for the Kurtz et al. (2004) survey.  In both
cases this is the result of the sample selection for the survey in
question.  The Slysh et al. survey was one of the first large
class I methanol maser searches in the southern sky and employed a
diverse source sample, in contrast the Kurtz et al. work targeted
many sources already known to contain 44 GHz class I methanol
masers.

We consider a class~I methanol maser to be associated with an EGO
if the separation of the maser and the EGO is less than 1\arcmin.
The positional accuracy of the {\em Spitzer} Galactic Legacy
Infrared Mid-Plane Survey Extraordinaire (GLIMPSE) point source
catalogue is better than 1\arcsec, however, EGOs are extended
objects \citep[angular extents between a few to $>$30\arcsec
;][]{C+08} and the positions of many of the class~I masers have
only been determined with single dishes and have positional
uncertainties of the order of 1\arcmin. Past experience with maser
searches targeted towards {\em IRAS} point sources suggests that
some of the associations we find will be simply due to EGOs and
class I masers both being found in high-mass star formation
regions rather than a physical association. This means that the
results of our investigation provide upper limits on the rate of
physical association between EGOs and class I methanol masers.

Table~2 shows the detection rate towards EGOs for each of the
class~I maser surveys in our sample, in each case the detection
rate exceeds 50\%. Combining the data from all surveys and both
transitions a total of 61 EGOs (including 41 likely outflow
sources and 20 possible candidates) have been searched, with
detections towards 41 sources (of which 28 lie in likely outflow
sources and 13 in possible outflow ones) and non-detections in the
other 20 cases - yielding a detection rate of 67\%. Information on
the 44- and 95-GHz methanol masers associated with EGOs (including
source name, position, association with 6.7-GHz methanol maser,
whether the source is detected at 44 and/or 95~GHz etc) is
summarised in Table~3.  The 6.7-GHz class~II methanol maser
information has been obtained from the catalogues of Walsh et al.
(1998), Pestalozzi et al. (2005), Ellingsen (2006) and Pandian et
al. (2007).

\citet{W+04} compiled a catalogue of high-velocity outflow
sources, the majority of which are bipolar outflows and 38\% of
which are associated with high-mass star formation regions. They
used either the bolometric luminosity of the central source,
L$_{bol}$, or the outflow gas mass, M, as criteria to distinguish
between high-mass sources and low-mass sources, (i.e. a source is
considered to be high-mass source if either L$_{bol}$ $>$ 1000
L$_{\odot}$ or M $>$ 3 M$_{\odot}$). We have also compared the
class~I methanol maser sample to the \citeauthor{W+04} catalogue,
after removing sources which are also EGOs. Using the same
association criterion (1\arcmin) we find that an additional 34
outflow sites (29 high-mass sources and 5 low-mass sources) have
been searched for class~I methanol masers, with detections towards
23 sources (of which 21 are high-mass outflows and 2 are low-mass
outflows) and non-detections towards the remaining 11 (Table~2).
This detection rate of 68\% that is the same (within the
statistical uncertainty) with that found for EGOs. Information on
the 44- and 95-GHz methanol masers associated with sources in this
outflow catalogue (including source name, position, association
with 6.7-GHz methanol maser, whether the source is detected at 44
and/or 95~GHz etc) is summarised in Table 4. The low-mass sources
are marked by ``$\ast$'' in this table. Note that there were no
class~I masers from the survey of Ellingsen (2005) associated with
outflow sources from the Wu et al. catalogue.

Combining the sample of class~I methanol masers associated with
EGOs and those associated with other outflow sources (Wu et al.
catalogue), means that in total 64 of the 135 class~I methanol
masers in our sample ($\sim$ 50\%) are associated with outflows,
identified either from infrared data or from millimeter line
observations. Table~2 summarises the detection rate for each of
the class~I maser surveys for EGOs, other outflows, it also
provides detection rates for the 44- and 95-GHz class I methanol
maser transitions individually and the total over all surveys. The
totals shown are not simply the sum of the individual surveys
because some of the sources appear in two or more of the surveys.

\section{Discussion} \label{sec:discussion}

Although the association between class~I methanol masers and
outflows was first proposed more than 20 years ago and there is
strong evidence for the association in a small number of sources,
statistical evidence for the relationship has been scarce with
\citet{VL07} finding less than 25\% of class~I masers were
associated with outflows within 2\arcmin. In contrast we have
found that approximately two-thirds of high-mass outflow sources
have an associated known class~I methanol maser and that
approximately 50\% of class~I masers are associated with an
outflow. The difference between 25\% and 50\% of the associations
is due to a larger variety of outflow sources used in our study
(EGOs and Wu et al. sample). The best targeting criterion which
had previously been developed to search for class~I methanol
masers was to search towards known class~II methanol masers, which
has been shown to yield a detection rate of approximately 40\%
\citep{E05}. \citet{C+08} identified 302 likely and possible EGOs
from the GLIMPSE I survey, many more EGOs are likely to be found
when a similar investigation is undertaken for the $|l| <
10^{\circ}$ region (GLIMPSE II). To date only 61 of the 302 EGOs
have been searched for class~I methanol maser emission, so if the
two-thirds detection rate holds across the whole sample we might
expect a targeted search towards EGOs to find approximately
another 160 class~I methanol masers, doubling the number of
sources currently known. A more detailed understanding of the
relationship between EGOs and class~I methanol masers may allow
them to be used to further refine the lists of outflow candidates.

Outflows are commonly observed in both low- and high-mass star
forming regions, and indeed the majority of known outflow sources
are associated with low-mass star formation \citep{W+04}. Most
known class~I methanol masers appear to be associated with
high-mass star forming regions, however, weak maser emission has
recently been detected towards a number of low- and
intermediate-mass star formation regions \citep{K+06}. Our
statistical analysis of the known outflow sources listed in Wu et
al. (2004) also shows that 21 class~I methanol masers are
associated with high-mass star forming regions and only 2 masers
are associated with low-mass ones (OMC-2 and Mol 138 listed in
Table 4; both sources are considered to be low-mass sources due to
their outflow gas mass M $<$ 3 M$_{\odot}$ on the basis of Wu et
al. criterion). \citet{C+08} argue that EGOs are only associated
with high-mass star formation. This is based on their association
with other high-mass star formation tracers such as IRDC and
class~II methanol masers, and that the surface brightness of
low-mass outflows will be too faint to be detected at the
sensitivity of the GLIMPSE survey. The detection of strong class~I
methanol masers towards a large fraction of the EGOs which have
been searched, is consistent with the results of \citeauthor{C+08}
that EGOs are associated with outflows from MYSOs.

The statistical results also show that approximately 50\% of known
class~I methanol masers are not associated with a known outflow
traced either by an EGO or millimeter thermal lines. There are
three possible reasons why this may be the case for an individual
source:
\begin{enumerate}

\item No observations for outflows have been made towards the
source (e.g. it lies outside the region covered by the GLIMPSE I
survey and has not been targeted for molecular line mapping).

\item It is associated with an outflow, but the observations which
have been made to date are not sensitive enough to detect the
outflow (most likely for distant sources).

\item There is no outflow associated with the source.

\end{enumerate}

The first two cases may be the correct explanation for most, or
even all of the current non-outflow associated sources. It is
important to target future outflow searches towards these class~I
methanol masers to determine if some sources are not associated
with outflows as this would provide new insights into the
mechanism through which class~I methanol masers are produced.

It must be remembered that the results of the statistical study we
have undertaken here may be influenced by target selection effects
in the four class I maser surveys we have investigated. The high
rate of association of currently known class I masers with class
II masers (72\% as reported by Val'tts \& Larionov (2007)) is an
example of one such possible selection effect. To date, the
majority of class I masers have been detected towards known class
II masers, for which a good association with EGOs has already been
demonstrated by Cyganowski et al. Table 3 clearly shows that most
EGOs (57/61) associated with class I masers are also associated
with 6.7-GHz class II masers within 1\arcmin. According to
Cyganowski et al. (2008) the majority of EGOs (73\% in the likely
sample) are associated with 6.7-GHz class II masers, and the
sample of known class I masers is largely a subset of the sample
of known class II masers, thus a high detection rate for class I
masers towards EGOs is not unexpected. However, Table 3 also shows
that there are EGOs (e.g. G343.12-0.06, G348.18+0.48) which have
class I masers but are not associated with class II masers
(actually these two sources don't have 6.7-GHz masers brighter
than the detection limit of the Parkes multi-beam survey Green et
al. 2009; Caswell et al. in prep.). This demonstrates that the
EGO-based selection is likely to be complementary to other
methods, especially to class II maser selection method. However, a
more definitive answer on the role of selection effects can only
be obtained through a class~I maser search towards a EGO-based
target sample or an untargeted class~I maser search.

\subsection{An evolutionary sequence for high-mass star formation}

The formation of stars of 8 M$_\odot$ and greater, remains an area
of hot debate in modern astrophysics, with competing theories as
to whether they form through scaled-up versions of the
accretion-driven process observed in low-mass star formation
regions, or through mergers of low- and intermediate- mass stars
at the center of dense stellar clusters \citep[see][for an
overview of the arguments]{BZ05}. Both outflows and interstellar
masers are frequently used tools for studying the high-mass star
formation process. Outflows are driven by infall and accretion
processes and so detection of an active outflow associated with an
O-star would provide strong evidence for accretion as the
mechanism for high-mass star formation, however, to date only a
few candidates have been found \citep[e.g.][]{Bro03}. Class II
methanol masers can be used to probe the kinematics and physical
conditions in high-mass star formation regions at milliarcsecond
resolutions and, in a number of sources it is suggested that the
masers arise in a disk \citep[e.g.][]{P+04,B+05}. However, the
existing data is not sufficient to allow a good determination of
parameters such as the disk mass and radius from the maser data.

In the medium-term, statistical studies of outflows and masers may
be of greater utility for high-mass star formation studies than
detailed investigations of specific sources. They are both good
signposts of the early stages of high-mass star formation and both
have been suggested as potential evolutionary tracers. \cite{BS05}
suggest that the degree of collimation of outflows decreases as
the mass of the star increases (that accretion is the mechanism
for high-mass star formation is implicit in this scenario).  So
more highly collimated outflows are associated with younger (and
hence lower-mass) massive stars. \cite{E+07} have also suggested
an evolutionary scheme for high-mass star formation traced by the
presence/absence of different maser species and transitions. In
this scheme it is proposed that class~I methanol masers may be the
first maser species to switch on during the process of high-mass
star formation. A better understanding of the relationship between
outflows traced by EGOs and class~I methanol masers, which we have
revealed, may allow these two evolutionary schemes to be compared
for consistency and perhaps merged into a single more
comprehensive and broadly based scenario. This could be achieved
through detailed high resolution studies of a sample of class~I
methanol masers and their comparison with mid-infrared data.

\section{Conclusion}

We have demonstrated that outflows traced by both EGOs and
millimeter molecular line observations provide reliable targeting
criteria (a detection rate of $\sim$ 67\%) for searching for
class~I methanol masers. This provides statistical confirmation
for the long-purported relationship between class~I methanol
masers and outflows and provides a large new sample of source
towards which searches can be targeted. Masers and outflows are
both frequently used as signposts of high-mass star formation and
a better understanding of the relationship between class~I
methanol masers and outflows promises to enable the development of
a more broadly-based evolutionary scheme for high-mass star
formation.

\section*{Acknowledgments}

We thank the referee for helpful comments that improved the
manuscript. This research has made use of NASA's Astrophysics Data
System Abstract Service. This research has made use of data
products from the GLIMPSE survey, which is a legacy science
program of the {\em Spitzer Space Telescope}, funded by the
National Aeronautics and Space Administration. This work was
supported in part by the National Natural Science Foundation of
China (grants 10803017, 10573029, 10625314, 10633010 and 10821302)
and the Knowledge Innovation Program of the Chinese Academy of
Sciences (Grant No. KJCX2-YW-T03), and the National Key Basic
Research Development Program of China (No. 2007CB815405).



\label{lastpage}

\begin{table*}
 \centering
 \begin{minipage}{200mm}
  \caption{Summary of major class I methanol maser surveys.}
  \begin{tabular}{@{}llrrrrlrlr@{}}
\hline
Survey & Transition & \multicolumn{2}{c}{\# Sources} & Approx. & Sample selection \\
 & & Searched & Detected &  Sensitivity (Jy) & \\
\hline
\citet{S+94}    & 44 GHz ($7_0$--$6_1$ A$^{+}$) & $\sim$250  & 59 & 5 -- 10   & H\,{\sevensize II} regions, water and 6.7 GHz methanol masers \\
\citet{V+00}    & 95 GHz ($8_0$--$7_1$ A$^{+}$) & 153        & 85 & $\sim 6$  & H\,{\sevensize II} regions, known methanol at 6.7, 36 or 44 GHz\\
\citet{KHA04}   & 44 GHz ($7_0$--$6_1$ A$^{+}$) & 44         & 37 & $< 0.3$   & Known 44 GHz, IR selected MYSO \\
\citet{E05}     & 95 GHz ($8_0$--$7_1$ A$^{+}$) & 60         & 26
& $\sim 6$  & 6.7 GHz methanol masers \\
\hline
\end{tabular}
\end{minipage}
\end{table*}


\begin{table*}
 \centering
 \begin{minipage}{120mm}
  \caption{Summary of Class~I methanol maser associations.}
  \begin{tabular}{lrrrcrrrrc}
\hline
  & \multicolumn{4}{c}{EGOs} &  \multicolumn{4}{c}{Other Outflows $^a$ }  \\
  & \multicolumn{4}{c}{Maser Detection} &  \multicolumn{4}{c}{Maser Detection}   \\
 Survey & Yes & No & Rate & Uncer.$^b$ & Yes & No & Rate & Uncer.$^b$ \\
\hline
\citet{S+94}    & 23 & 19 & 55\% & 15\% &  4 &  7 & 36\% & 30\% \\
\citet{V+00}    & 28 &   6 & 82\% & 17\% & 11 & 4 & 73\% & 26\% \\
\citet{KHA04}   &   4 &   0 & 100\% & 50\% & 15 & 3 & 83\% & 24\% \\
\citet{E05}     &  13 & 5 & 72\% & 24\% & ... & ... & ... & ... $^c$ \\
\hline
Total $^d$   &       &     &           &   &    &      \\
44~GHz          & 27 & 19 & 59\% & 15\% & 18 & 10 & 64\% & 19\%  \\
95~GHz          & 34 & 11 & 76\% & 15\% & 11 &   4 & 73\% & 26\% \\
Class~I $^e$ & 41 & 20 & 67\% & 13\% & 23 & 11 & 68\% & 17\% \\
\hline
\\
\end{tabular}

$^a$ The outflow sources in the catalogue of Wu et al. (2004)
after removing EGOs.

$^b$ Uncertainties in the detection rates are estimated from the
reciprocals of the square root

\ \ \ of source number for each case.

$^c$ No sources are associated with other outflows in the survey
of Ellingsen (2005).

$^d$ Some sources appear in more than one of the four survey paper
and so the total sum is less

\ \ \ than the sum of the totals for the four papers.

$^e$ A source detected at either 44 or 95~GHz is counted here, but
only once in each case.

\end{minipage}
\end{table*}

\begin{table*}
\renewcommand{\arraystretch}{0.90}
 \centering
 \begin{minipage}{200mm}
  \caption{Class I and class II methanol masers towards EGOs.}
     \begin{tabular}{lccc@{}ccc@{}cclc}
 \hline
 & \multicolumn{2}{c}{Source Position $^a$}& & \multicolumn{2}{c} {Class II $^b$} & & \multicolumn{3}{c} {Class I $^b$} & \\

\cline{2-3} \cline{5-6} \cline{8-10} Source name & R.A. (2000)& Dec. (2000)&  &
6.7 GHz ?  & Cat. $^c$ & & 44 GHz ?& 95 GHz ? & Maser survey $^d$ & Remark $^e$ \\
 &  $\emph{h m s}$& $^{\circ}$ $ \arcmin $ $ \arcsec$ & & & & \\
\hline
G298.26+0.74    &12  11  47.7    &-61 46  21 &  & Y  & P05&  & N   &  -- &   S        &   1  \\
G305.80-0.24    &13  16  43.4    &-62 58  29 &  & Y  & P05&  & N   &  -- &   S        &   4  \\
G309.38-0.13    &13  47  23.9    &-62 18  12 &  & Y  & P05&  & --  &  Y  &   V        &   1  \\
G318.05+0.09    &14  53  42.6    &-59 08  49 &  & Y  & P05&  & Y   &  N  &   S, V     &   2  \\
G320.23-0.28    &15  09  52.6    &-58 25  36 &  & Y  & P05&  & N   &  -- &   S        &   2  \\
G323.74-0.26    &15  31  45.5    &-56 30  50 &  & Y  & P05&  & Y   &  Y  &   S, V     &   4  \\
G324.72+0.34    &15  34  57.5    &-55 27  26 &  & Y  & P05&  & Y   &  Y  &   S, V     &   1  \\
G326.48+0.70    &15  43  17.5    &-54 07  11 &  & Y  & E06&  & Y   &  Y  &   S, V     &   2  \\
G327.12+0.51    &15  47  32.7    &-53 52  39 &  & Y  & P05&  & N   &  N  &   S, E     &   1  \\
G326.78-0.24    &15  48  55.2    &-54 40  37 &  & N  & E06&  & N   &  -- &   S        &   1  \\
G327.40+0.44    &15  49  19.3    &-53 45  10 &  & Y  & E06&  & N   &  N  &   S, E     &   1  \\
G327.39+0.20    &15  50  18.5    &-53 57  07 &  & Y  & E06&  & --  &  Y  &   V        &   1  \\
G326.86-0.67    &15  51  13.6    &-54 58  03 &  & Y  & E06&  & --  &  Y  &   V        &   1  \\
G327.30-0.58    &15  53  11.2    &-54 36  48 &  & Y  & P05&  & Y   &  Y  &   S, V     &   3  \\
G328.81+0.63    &15  55  48.4    &-52 43  06 &  & Y  & E06&  & Y   &  Y  &   S, V, E  &   4  \\
G328.25-0.53    &15  57  59.7    &-53 58  00 &  & Y  & E06&  & Y   &  Y  &   S, V, E  &   2  \\
G329.47+0.50    &15  59  41.0    &-52 23  28 &  & Y  & E06&  & --  &  Y  &   V        &   2  \\
G329.03-0.20    &16  00  30.6    &-53 12  34 &  & Y  & E06&  & Y   &  Y  &   S, V, E  &   2  \\
G329.07-0.31    &16  01  09.9    &-53 16  02 &  & Y  & E06&  & --  &  Y  &   V, E     &   3  \\
G329.18-0.31    &16  01  47.4    &-53 11  44 &  & Y  & E06&  & --  &  Y  &   V        &   1  \\
G329.61+0.11    &16  02  03.1    &-52 35  33 &  & Y  & E06&  & --  &  N  &   V        &   1  \\
G329.41-0.46    &16  03  32.4    &-53 09  26 &  & Y  & E06&  & N   &  N  &   S, E     &   2  \\
G330.95-0.18    &16  09  52.7    &-51 54  56 &  & Y  & E06&  & N   &  N  &   S, E     &   4  \\
G330.88-0.37    &16  10  19.9    &-52 06  13 &  & Y  & P05&  & N   &  -- &   S        &   2  \\
G331.13-0.24    &16  10  59.8    &-51 50  19 &  & Y  & E06&  & Y   &  Y  &   S, V, E  &   2  \\
G331.34-0.35    &16  12  26.4    &-51 46  17 &  & Y  & E06&  & Y   &  Y  &   S, V, E  &   4  \\
G332.29-0.09    &16  15  45.2    &-50 55  52 &  & Y  & E06&  & --  &  Y  &   V, E     &   4  \\
G332.56-0.15    &16  17  12.1    &-50 47  14 &  & Y  & E06&  & --  &  N  &   V        &   1  \\
G332.60-0.17    &16  17  29.4    &-50 46  13 &  & Y  & E06&  & --  &  Y  &   V        &   2  \\
G332.35-0.44    &16  17  31.4    &-51 08  22 &  & Y  & E06&  & --  &  N  &   V        &   4  \\
G333.18-0.09    &16  19  45.6    &-50 18  34 &  & Y  & E06&  & N   &  Y  &   S, E     &   1  \\
G332.73-0.62    &16  20  02.8    &-51 00  32 &  & Y  & P05&  & N   &  -- &   S        &   2  \\
G332.94-0.69    &16  21  18.9    &-50 54  10 &  & Y  & E06&  & --  &  Y  &   E        &   1  \\
G333.47-0.16    &16  21  20.2    &-50 09  50 &  & Y  & E06&  & N   &  Y  &   S, E     &   2  \\
G332.96-0.68    &16  21  22.9    &-50 52  58 &  & Y  & E06&  & --  &  Y  &   E        &   1  \\
G333.13-0.56    &16  21  36.1    &-50 40  49 &  & Y  & E06&  & --  &  Y  &   E        &   4  \\
G335.06-0.43    &16  29  23.1    &-49 12  28 &  & Y  & E06&  & --  &  Y  &   E        &   2  \\
G335.79+0.18    &16  29  47.1    &-48 15  47 &  & Y  & P05&  & Y   &  N  &   S, V     &   2  \\
G335.59-0.29    &16  30  58.5    &-48 43  51 &  & Y  & P05&  & Y   &  Y  &   S, V     &   1  \\
G337.40-0.40    &16  38  50.4    &-47 28  04 &  & Y  & P05&  & Y   &  Y  &   S, V     &   4  \\
G338.92+0.55    &16  40  33.6    &-45 41  44 &  & Y  & P05&  & Y   &  Y  &   S, V     &   4  \\
G337.91-0.48    &16  41  10.3    &-47 08  06 &  & Y  & P05&  & Y   &  Y  &   S, V     &   2  \\
G340.06-0.23    &16  48  09.7    &-45 20  58 &  & N  & W  &  & N   &  -- &   S        &   4  \\
G340.78-0.10    &16  50  14.7    &-44 42  31 &  & Y  & P05&  & N   &  -- &   S        &   3  \\
G343.12-0.06    &16  58  16.6    &-42 52  04 &  & N  & C  &  & Y   &  Y  &   S, V     &   1  \\
G344.58-0.02    &17  02  57.7    &-41 41  54 &  & Y  & P05&  & N   &  -- &   S        &   1  \\
G344.23-0.57    &17  04  07.1    &-42 18  42 &  & Y  & P05&  & Y   &  Y  &   S, V     &   2  \\
G345.51+0.35    &17  04  24.6    &-40 43  57 &  & Y  & P05&  & Y   &  Y  &   S, V     &   5  \\
G345.00-0.22    &17  05  11.2    &-41 29  03 &  & Y  & P05&  & Y   &  Y  &   S, V     &   4  \\
G348.18+0.48    &17  12  08.0    &-38 30  52 &  & N  & C  &  & Y   &  N  &   S, V     &   4  \\
G348.73-1.04    &17  20  06.5    &-38 57  08 &  & Y  & P05&  & N   &  -- &   S        &   4  \\
G11.92-0.61     &18  13  58.1    &-18 54  17 &  & Y  & P05&  & Y   &  -- &   K        &   1  \\
G14.33-0.64     &18  18  54.4    &-16 47  46 &  & Y  & P05&  & Y   &  Y  &   S, V     &   1  \\
G16.59-0.05     &18  21  09.1    &-14 31  48 &  & Y  & P05&  & Y   &  Y  &   S, V     &   2  \\
G20.24+0.07     &18  27  44.6    &-11 14  54 &  & Y  & P05&  & N   &  -- &   S        &   4  \\
G23.01-0.41     &18  34  40.2    &-09 00  38 &  & Y  & P05&  & Y   &  Y  &   S, V     &   1  \\
G24.33+0.14     &18  35  08.1    &-07 35  04 &  & Y  & P05&  & N   &  -- &   S        &   4  \\
G28.83-0.25     &18  44  51.3    &-03 45  48 &  & Y  & P05&  & N   &  N  &   S, E     &   1  \\
G34.26+0.15     &18  53  16.4    &+01 15  07 &  & Y  & P05&  & Y   & --  &   K        &   5  \\
G43.04-0.45     &19  11  38.9    &+08 46  39 &  & Y  & P07&  & Y   &  -- &   K        &   4  \\
G45.47+0.05     &19  14  25.6    &+11 09  28 &  & Y  & P07&  & Y   &  -- &   K        &   1  \\
\hline
\\
\end{tabular}
\newline
$^a$ The coordinates are given by the EGO positions.

$^b$ Association with 6.7-GHz class II and 44, 95-GHz class I
methanol masers within 1\arcmin:

\ \ \ Y = Yes, N = No, ¡°-¡± = no information.

$^c$ The catalogue of 6.7-GHz class II methanol maser: P05 --
Pestalozzi et al. 2005, E06 -- Ellingsen 2006, P07 -- Pandian et
al. 2007,

\ \ \ C -- Caswell in prep., W -- Walsh et al. 1998.

$^d$ The survey work for class I methanol maser: E -- Ellingsen
2005, K -- Kurtz et al. 2004, S -- Slysh et al. 1994, V -- Val'tts
et al. 2000.

$^e$ 1, 2, 3, 4 and 5 represent the sources are selected from
Tables 1, 2, 3, 4 and 5 of Cyganowski et al. (2008), respectively.

\end{minipage}
\end{table*}

\begin{table*}
 \centering
 \begin{minipage}{200mm}
  \caption{Class I and class II masers towards outflow sources from
Wu et al. catalogue which are not EGOs.}
 \begin{tabular}{llccc@{}ccc@{}cclc}
 \hline
 & & \multicolumn{2}{c}{Source Position $^a$}& & \multicolumn{2}{c} {Class II $^b$} & & \multicolumn{3}{c} {Class I $^b$} & \\

\cline{3-4} \cline{6-7} \cline{9-11}

Source name & Other name & R.A. (2000)& Dec. (2000)&  &
6.7 GHz ?  & Cat. $^c$ & & 44 GHz ?& 95 GHz ? & Maser survey $^d$ \\
 & & $\emph{h m s}$& $^{\circ}$ $ \arcmin $ $ \arcsec$ & & & & \\
\hline

G139.91+0.20   & AFGL437          & 03 07 24.4  & +58 31 08  & &  -- & -- & & N  & --  & K    \\
G176.23-20.89* & T Tau            & 04 21 59.2  & +19 32 06  & &  -- & -- & & N  & N   & S, V    \\
G208.82-19.24* & OMC-2            & 05 35 27.7  & -05 09 40  & &  -- & -- & & -- & Y   & V    \\
G210.44-19.76* & CS-star HHI      & 05 36 20.8  & -06 45 35  & &  -- & -- & & N  & N   & S, V \\
G173.72+2.69   & S235B            & 05 40 52.5  & +35 41 26  & &  -- & -- & & Y  & --  & K    \\
G192.60-0.05   & S255             & 06 12 54.4  & +17 59 25  & &  Y  & P05& & Y  & --  & K    \\
G203.32+2.06   & NGC 2264         & 06 41 10.8  & +09 29 07  & &  -- & -- & & -- & Y   & V    \\
G263.25+0.52   & IRAS 08470-4243  & 08 48 48.2  & -42 54 25  & &  Y  & P05& & N  & N   & S, V \\
G274.01-1.15   & WB89 1275        & 09 24 26.0  & -51 59 34  & &  -- & --    & & N  & --  & S    \\
G301.13-0.22   & IRAS 12326-6245  & 12 35 34.1  & -63 02 28  & &  Y  & P05& & Y  & Y   & S, V \\
G302.03-0.06   & IRAS 12405-6238  & 12 43 32.5  & -62 55 12  & &  Y  & P05& & N  & --  & S    \\
G351.16+0.69   & NGC 6334         & 17 19 58.4  & -35 57 50  & &  Y  & P05& & Y  & Y   & S, V \\
G351.42+0.64   & NGC 6334I        & 17 20 55.1  & -35 46 44  & &  Y  & P05& & Y  & Y   & S, V, K \\
G5.89-0.39     & IRAS 17574-2403  & 18 00 30.6  & -24 03 58  & &  Y  & P05& & Y  & Y   & V, K  \\
G6.05-1.45     & M8E              & 18 04 53.2  & -24 26 42  & &  -- & --   &  & Y  & Y & S, V  \\
G9.62+0.19     & IRAS 18032-2032  & 18 06 14.8  & -20 31 39  & &  Y  & P05& & Y  & --  & K    \\
G18.34+1.77    & Mol 146          & 18 17 57.1  & -12 07 22  & &  Y  & P05& & -- & Y   & V    \\
G10.84-2.59    & GGD 27           & 18 19 12.1  & -20 47 26  & &  -- & -- & & Y  & Y   & V, K \\
G17.64+0.15    & CRL2136          & 18 22 26.8  & -13 30 15  & &  Y  & P05& & N  & --  & S    \\
G16.87-2.16    & L379IRS3         & 18 29 24.9  & -15 15 49  & & --  & P05& & -- & Y   & V    \\
G25.41+0.11    & IRAS 18345-0641  & 18 37 16.5  & -06 38 32  & &  Y  & P05& & -- & N   & V    \\
G40.50+2.54    & IRAS 18537+0749  & 18 56 10.8  & +07 53 28  & &  -- & -- & & -- & Y   & V    \\
G359.93-17.85* & R CrA            & 19 01 54.4  & -36 57 10  & &  -- & -- & & N  & --  & S    \\
G45.07+0.13    & IRAS 19110+1045  & 19 13 21.7  & +10 50 53  & &  Y  & P05& & Y  & --  & K    \\
G45.12+0.13    &                  & 19 13 28.6  & +10 53 22  & &  -- & -- & & N  & --  & K    \\
G78.12+3.63    & Mol 119          & 20 14 26.0  & +41 13 32  & &  Y  & P05& & Y  & --  & K    \\
G75.78+0.34    & ON2              & 20 21 44.1  & +37 26 42  & &  Y  & P05& & Y  & --  & K    \\
G81.88+0.78    & W75N             & 20 38 37.4  & +42 37 57  & &  Y  & P05& & Y  & --  & K    \\
G81.68+0.54    & DR21             & 20 39 00.0  & +42 19 28  & &  -- & -- & & Y  & --  & K    \\
G94.26-0.41    & Mol 136          & 21 32 31.2  & +51 02 22  & &  -- & -- & & Y  & --  & K    \\
G99.98+4.17*   & Mol 138          & 21 40 42.1  & +58 16 10  & &  -- & -- & & Y  & --  & K    \\
G111.54+0.78   & NGC 7538         & 23 13 44.7  & +61 28 10  & &  Y  & P05& & Y  & --  & K    \\
G111.87+0.82   & Mol 155          & 23 16 11.8  & +61 37 45  & &  -- & -- & & N  & --  & K    \\
G114.53-0.54   & Mol 160          & 23 40 51.1  & +61 10 29  & &  -- & -- & & Y  & --  & K    \\
\hline
\\
\end{tabular}
\newline
$^a$ The coordinates are given by the outflow positions listed in
Wu et al. (2004).

$^b$ Association with a 6.7-GHz class II and 44, 95-GHz class I
methanol masers within 1\arcmin:

\ \ \ Y = Yes, N = No, ¡°-¡± = no information.

$^c$ The catalogue of 6.7-GHz class II methanol maser: P05 --
Pestalozzi et al. 2005.

$^d$ The survey work for class I methanol maser: K -- Kurtz et al.
2004, S -- Slysh et al. 1994, V -- Val'tts et al. 2000.

* Low-mass outflow source identified from Wu et al. (2004).

\end{minipage}
\end{table*}


\begin{thebibliography}{99}

\bibitem[Bally \& Zinnecker(2005)]{BZ05}
  Bally J., Zinnecker H., 2005, AJ, 129, 2281

\bibitem[Bartkiewicz et al.(2005)]{B+05}
  Bartkiewicz A., Szymczak M., van Langevelde H.\ J., 2005, A\&A, 442, L61

\bibitem[Batrla \& Menten(1988)]{BM88}
  Batrla W., Menten K.\ M., 1988, ApJ, 329, L117

\bibitem[Beuther \& Shepherd(2005)]{BS05}
  Beuther H., Shepherd D., 2005, in Cores to Clusters: Star Formation with Next Generation Telescopes, ed. N.\ Kumar, M.\ Tafalla, P.\ Caselli (New York: Springer), 105

\bibitem[Beuther et al.(2002)]{B+02}
  Beuther H., Schilke P., Gueth F., McCaughrean M., Andersen M., Sridharan T. K., Menten K. M., 2002, A\&A, 387, 931

\bibitem[Brooks et al.(2003)]{Bro03}
  Brooks K.\ J., Garay G., Mardones D., Bronfman L., 2003, ApJ,
  594, L131

\bibitem[Cragg et al.(1992)]{C+92}
  Cragg D.\ M., Johns K.\ P., Godfrey P.\ D., Brown R.\ D., 1992, MNRAS, 259, 203

\bibitem[Cyganowski et al.(2008)]{C+08}
  Cyganowski C.\ J. et al., 2008, AJ, 136, 2391

\bibitem[De Buizer et al.(2009)]{dB+09}
  De Buizer J.\ M., Redman R.\ O., Longmore S.\ N., Caswell J.\ L., Feldman P.\ A., 2009, A\&A, 493, 127

\bibitem[Green et al.(2009)]{G+09}
  Green J.\ A. et al., 2009, MNRAS, 392, 783

  \bibitem[Ellingsen(2005)]{E05}
   Ellingsen S.\ P., 2005, MNRAS, 359, 1498

 \bibitem[Ellingsen(2006)]{E06}
   Ellingsen S.\ P., 2006, ApJ, 638, 241


\bibitem[Ellingsen et al.(2007)]{E+07}
  Ellingsen S.\ P., Voronkov M.\ A., Cragg D.\ M., Sobolev A.\ M., Breen S.\ L., Godfrey P.\ D., 2007 in Astrophysical Masers and their Environments, Proceedings of IAU Symposium 242 (Cambridge University Press), 217

 \bibitem[Johnston et al.(1997)]{J+97}
   Johnston K.\ J., Gaume R.\ A., Wilson T.\ L., Nguyen H.\ A., Nedoluha G.\ E., 1997, ApJ, 490, 758

 \bibitem[Kalenskii et al.(2006)]{K+06}
   Kalenskii S.\ V., Promyslov V.\ G., Slysh V.\ I., Bergman P., Winnberg A., 2006, Astron. Rep., 50, 289

 \bibitem[Kurtz et al.(2004)]{KHA04}
   Kurtz S., Hofner P., \'Alvarez C.V., 2004, ApJS, 155, 149

\bibitem[Menten(1991)]{M91}
  Menten K.~M., 1991 in ``Skylines'' proceedings of the Third Haystack Observatory Meeting, ed. A.\ D.\ Haschick, P.\ T.\ P.\ Ho (San Fransisco: Astronomical Society of the Pacific), 119

\bibitem[Minier et al.(2003)]{MENB03}
  Minier V., Ellingsen S.\ P., Norris R.\ P., Booth R.\ S., 2003, A\&A, 403, 1095

\bibitem[Pandian et al. (2007)] {Pan07}
Pandian J.\ D., Goldsmith P.\ F., Deshpande A. A., 2007, ApJ, 656,
255

\bibitem[Pestalozzi et al.(2002)]{P+02}
  Pestalozzi M.\ R., Humphreys E.\ M.\ L., Booth R.\ S., 2002, A\&A, 384, L15

\bibitem[Pestalozzi et al.(2004)]{P+04}
  Pestalozzi M.\ R., Elitzur M., Conway J.\ E., Booth R.\ S., 2004, ApJ, 603,
  113

\bibitem[Pestalozzi et al.(2005)]{PMB05}
  Pestalozzi M.\ R., Minier V., Booth R.\ S., 2005, A\&A, 432, 737

\bibitem[Plambeck \& Menten(1990)]{PM90}
  Plambeck R.\ L., Menten K.\ M., 1990, ApJ, 364, 555

\bibitem[Qiu et al.(2008)]{Q+08}
  Qiu K. et al., 2008, ApJ, 685, 1005

\bibitem[Slysh et al.(1994)]{S+94}
  Slysh V.\ I., Kalenskii S.\ V., Val'tts I.\ E., Otrupcek R., 1994, MNRAS, 268, 464

 \bibitem[Val'tts et al.(2000)]{V+00}
   Val'tts I.\ E., Ellingsen S.\ P., Slysh V.\ I., Kalenskii S.\ V., Otrupcek R., Larinov G.\ M., 2000, MNRAS, 317, 315

\bibitem[Val'tts \& Larinov(2007)]{VL07}
  Val'tts I.\ E., Larinov G.\ M., 2007, Astron. Rep., 51, 519

\bibitem[Voronkov et al.(2005)]{V+05}
  Voronkov M.\ A., Sobolev A.\ M., Ellingsen S.\ P., Ostrovskii A.\ B., 2005, MNRAS, 362, 995

\bibitem[Voronkov et al.(2006)]{V+06}
  Voronkov M.\ A., Brooks K.\ J., Sobolev A.\ M., Ellingsen S.\ P., Ostrovskii A.\ B., Caswell J.\ L., 2006, MNRAS, 373, 411

\bibitem[Walsh et al.(1998)]{W+98}
  Walsh A.\ J., Burton M.\ G., Hyland A.\ R., Robinson\ G., 1998, MNRAS, 301, 640

\bibitem[Walsh et al.(2003)]{W+03}
  Walsh A.\ J., Macdonald G.\ H., Alvey N.\ D.\ S., Burton M.\ G., Lee J.\ -K., 2003, A\&A, 410, 597


\bibitem[Wu et al.(2004)]{W+04}
  Wu Y., Wei Y., Zhao M., Shi Y., Yu W., Qin S., Huang M., 2004, A\&A, 426, 503


\end{thebibliography}
\end{document}